\documentclass[%
 aip,
 jmp,%
 amsmath,amssymb,
 reprint,%
]{revtex4-1}
\usepackage{multirow,color}
\usepackage{graphicx}
\usepackage{dcolumn}
\usepackage{bm}
\usepackage{etoolbox}
\newcommand{\m}{$\mu$}
\begin{document}

\preprint{AIP/123-QED}

\title[HBAR gravimetric sensitivity: towards wideband acoustic spectroscopy
]{High-overtone Bulk-Acoustic Resonator gravimetric sensitivity: towards wideband 
acoustic spectroscopy}

\author{D. Rabus}
\altaffiliation{FEMTO-ST, UMR CNRS-UFC-ENSMM-UTBM 6174, ENSMM, 26 Chemin de 
	l'\'Epitaphe, 25030 Besan\c con Cedex, France}
\altaffiliation{SENSeOR, Temis Innovation, 18 rue Alain Savary, 25000 
	Besan\c con, France}

\author{J.M. Friedt}
\altaffiliation{FEMTO-ST, UMR CNRS-UFC-ENSMM-UTBM 6174, ENSMM, 26 Chemin de 
	l'\'Epitaphe, 25030 Besan\c con Cedex, France}
\altaffiliation{SENSeOR, Temis Innovation, 18 rue Alain Savary, 25000 
	Besan\c con, France}

\author{S. Ballandras}
\altaffiliation{Frec$|$n$|$sys, Temis Innovation, 18 rue Alain Savary, 25000 
Besan\c con, France}

\author{T. Baron}
\altaffiliation{FEMTO-ST, UMR CNRS-UFC-ENSMM-UTBM 6174, ENSMM, 26 Chemin de 
	l'\'Epitaphe, 25030 Besan\c con Cedex, France}

\author{\'E. Lebrasseur}
\altaffiliation{FEMTO-ST, UMR CNRS-UFC-ENSMM-UTBM 6174, ENSMM, 26 Chemin de 
	l'\'Epitaphe, 25030 Besan\c con Cedex, France}
\author{\'E. Carry}
\altaffiliation{FEMTO-ST, UMR CNRS-UFC-ENSMM-UTBM 6174, ENSMM, 26 Chemin de 
	l'\'Epitaphe, 25030 Besan\c con Cedex, France}

\date{\today}

\begin{abstract}
In the context of direct detection sensors with compact dimensions, we
investigate the gravimetric sensitivity of High-overtone Bulk Acoustic
Resonators, through modeling of their acoustic characteristics and experiment.
The high frequency characterizing such devices is expected to induce a significant
effect when the acoustic field boundary conditions are modified by a thin adlayer.
Furthermore, the multimode spectral characteristics is considered for
wideband acoustic spectroscopy of the adlayer, once the gravimetric sensitivity
dependence of the various overtones is established. Finally, means of
improving the gravimetric sensitivity by confining the acoustic field in a low
acoustic-impedance layer is theoretically established.
\end{abstract}

\maketitle

\section{Introduction}

Direct detection sensors \cite{gizeli2004biomolecular} aim at continuous, real time monitoring of the 
presence and concentration of chemical compounds without the need of a preliminary
sample preparation step. Amongst the various direct detection strategies including
electrochemical methods \cite{thevenot2001electrochemical,janata2009principles}, 
optical methods (surface plasmon resonance, integrated
optics or spectroscopies \cite{homola1999surface,taules2012overview}), 
the use of acoustic waves to probe medium property variations 
is considered in contexts in which other strategies are not suitable either due to
the fragile optical setups or because the compound being investigated is not electrochemically
active. The two broad strategies of acoustic transducers aim at observing either boundary
condition variations due to the absorption of a thin film on a substrate in which the
acoustic wave is confined (the so-called Quartz Crystal Microbalance -- QCM\cite{su2005qcm,si2007polymer}), 
or acoustic
velocity variations as the boudary conditions are varied by chemical absorption on the surface
of the transducer guiding the propagation of a wave confined to the piezoelectric transducer
surface (the so-called Surface Acoustic Wave -- SAW). Various wave polarization conditions meet
the surface confinement requirements but only pure shear waves and waves exhibiting acoustic velocities
slower than those of the surrounding medium will prevent radiation losses as the sensor is loaded
by a liquid: the former approach is implemented in the Love mode transducer concept 
\cite{moll2007love,tamarin2003study} and the latter in the Lamb wave transducer. All these strategies
have been thoroughly investigated in the context of direct detection (bio)sensors. The evolution
from the QCM to the SAW strategy has been motivated by the consideration that rising acoustic
frequencies lowers the acoustic wavelength and hence magnifies the effect of a chemical species
absorption to form a layer of a given thickness: the gravimetric sensitivity quantifies this
notion. Rising QCM frequency classically means lowering the substrate thickness and hence 
making the transducer more fragile. An alternative consideration here is to use a
thin piezoelectric film over a thick substrate selected for its low acoustic losses to both
provide high acoustic frequency modes and yet a rugged transducer.

The work presented here focuses on the study of the gravimetric sensitivity
by modeling the acoustic transducer electrical response with a one dimensional model. The 
dependency of  the gravimetric sensitivity on the working frequency is demonstrated. The 
influence of the adsorbed thickness of the added layer and its acoustic properties on 
the gravimetric sensitivity is also presented. A discussion is proposed on the gravimetric 
sensitivity definition which depends on the considered initial condition.
A maximum of sensitivity is 
also obtained for a particular thickness in function of acoustic wavelength. 
The theoretical results are compared with experimental results obtained by considering
copper thin film deposition in dry and wet environments. Finally a way to improve the gravimetric 
sensitivity is proposed using an appropriate added layer on the sensing surface of the transducer.

\section{Acoustic wave transducers}

The High-overtone Bulk-Acoustic Resonator (HBAR) concept has evolved from the bulk-acoustic
resonator (QCM) strategy by identifying a technological limitation to how thin
a piezoelectric film could be made when aiming at rising operating frequencies $f_0$
\cite{abe2012fabrication,xeco::site}.
Since a QCM confines half a wavelength $\lambda$ in the substrate thickness $t$,
the resonator frequency is related to the acoustic velocity $v$ by $f_0=c/\lambda=v/(2t)$:
reaching low $t$ values has been investigated in the free membrane strategy of 
the Film Bulk Acoustic Resonator (FBAR) \cite{nirschl2009film,xu2011high}. 
The HBAR prevents the fragile piezoelectric membrane
from collapsing by being supported on a low acoustic loss substrate.

This work focuses on the determination of gravimetric sensitivity 
(Eq. \ref{sensi1}) of HBAR to assess the possibility of using such a transducer
for direct detection, and various sensing strategies introduced by the
unique spectral properties of the device. Assuming a linear 
relation between an adsorbed mass $\Delta m$ and the transducer resonance frequency 
shift $\Delta f$, the gravimetric sensitivity $S$ is defined as the relative 
frequency shift $\frac{\Delta f}{f_0}$ of the resonance when loading the
sensing area $A$
\begin{equation}
S=\frac{\Delta f}{f_0} \times \frac{A}{\Delta m}=\frac{\Delta f}{f_0} \times 
\frac{1}{\rho \times \Delta t}
\label{sensi1}
\end{equation}
since $\Delta m= A\rho\Delta t$ with $\rho$ the absorbed layer density and $\Delta t$ its thickness.

Eq.  \ref{sensi1} is used throughout this work for computing $S$ out of the modeled acoustic
transducer frequency variation due to layers with various properties being added over the
transducer surface. {\color{red}However, another practical quantity relating directly frequency shift and 
absorbed mass is the mass-sensitivity constant $C=\frac{\Delta m}{A\cdot \Delta f}$ in ng.cm$^{-2}$.Hz$^{-1}$:
the relatioship between these two quantities is $C=\frac{10^9}{S\cdot f_0}$.}

The perturbative approach of Sauerbrey \cite{sauerbrey} predicts (Eq. \ref{Sauerbrey2}) a 
gravimetric sensitivity only dependent on the transducer thickness $t_p$ and the density 
of the piezoelectric material $\rho_p$, assuming the adsorbed
layer is characterized by $\rho=\rho_p$. Hence, a perturbative model 
hints at a lack of improvement of the gravimetric sensitivity when using high overtone devices which 
are expected to always exhibit the fundamental mode gravimetric sensitivity. 
\begin{equation}
S= \frac{1}{\rho_p\cdot t_p}
\label{Sauerbrey2}
\end{equation}
which results from considering, in a perturbative approach, that
$\frac{\Delta f}{f}=\frac{\Delta \lambda}{\lambda}$ and that the wavelength 
$\lambda_n$ of the $n$th overtone is related to the substrate thickness by 
$t_p=\frac{n\lambda_n}{2}$, so that Eq. \ref{sensi1} can be 
written for each overtone of the QCM as
\begin{equation}
S=\frac{\Delta f_n}{f_n} \times \frac{2}{\rho_p n \Delta 
\lambda_n} = 
\frac{\Delta
	\lambda_n}{\lambda_n} \times \frac{2}{\rho_p n \Delta 
	\lambda_n}
 = \frac{2}{\rho_p \times \lambda_1}
\label{Sauerbrey_n}
\end{equation}
with $\lambda_1=n\cdot \lambda_n=t_p/2$ the wavelength of the fundamental mode.

Numerical modeling will however be considered to finely analyze the gravimetric sensitivity 
of HBAR overtones beyond these perturbative assumptions, if only because the HBAR is a complex
structure yielding more complex behaviours than this expected constant gravimetric sensitivity
with overtone number. {\color{red}Once the sensitivity is established, the detection limit
for a resonator operating at frequency $f_0$ and exhibiting a quality factor $Q$ is given
by the phase to frequency slope $\frac{d\varphi}{df}=\frac{2Q}{f_0}$. From the sensitivity
definition, knowing the smallest detectable phase shift $d\varphi_{min}$ as given in our case in
\cite{rabus2013high}, then the smallest relative detectable frequency shift is 
$\frac{df_{min}}{f_0}=\frac{d\varphi_{min}}{2Q}$ and 
$S=\frac{df}{f_0}\times\frac{A}{dm}\Leftrightarrow 
\frac{dm}{A}=\frac{1}{S}\times \frac{df}{f_0}\Rightarrow \frac{dm_{min}}{A}=\frac{d\varphi_{min}}{2Q}\times\frac{1}{S}$. 
As an example of a numerical application, considering a quality factor of $Q=10000$ and a minimum detectable
phase variation \cite{rabus2013high} of $d\varphi_{min}=25\mbox{~m}^\circ$ and $S=150$~cm$^2$/g,
then $\frac{dm_{min}}{A}\simeq 4$~ng/cm$^2$.}

\begin{figure}[h!tb]
	\begin{center}
		\includegraphics[width=\linewidth]{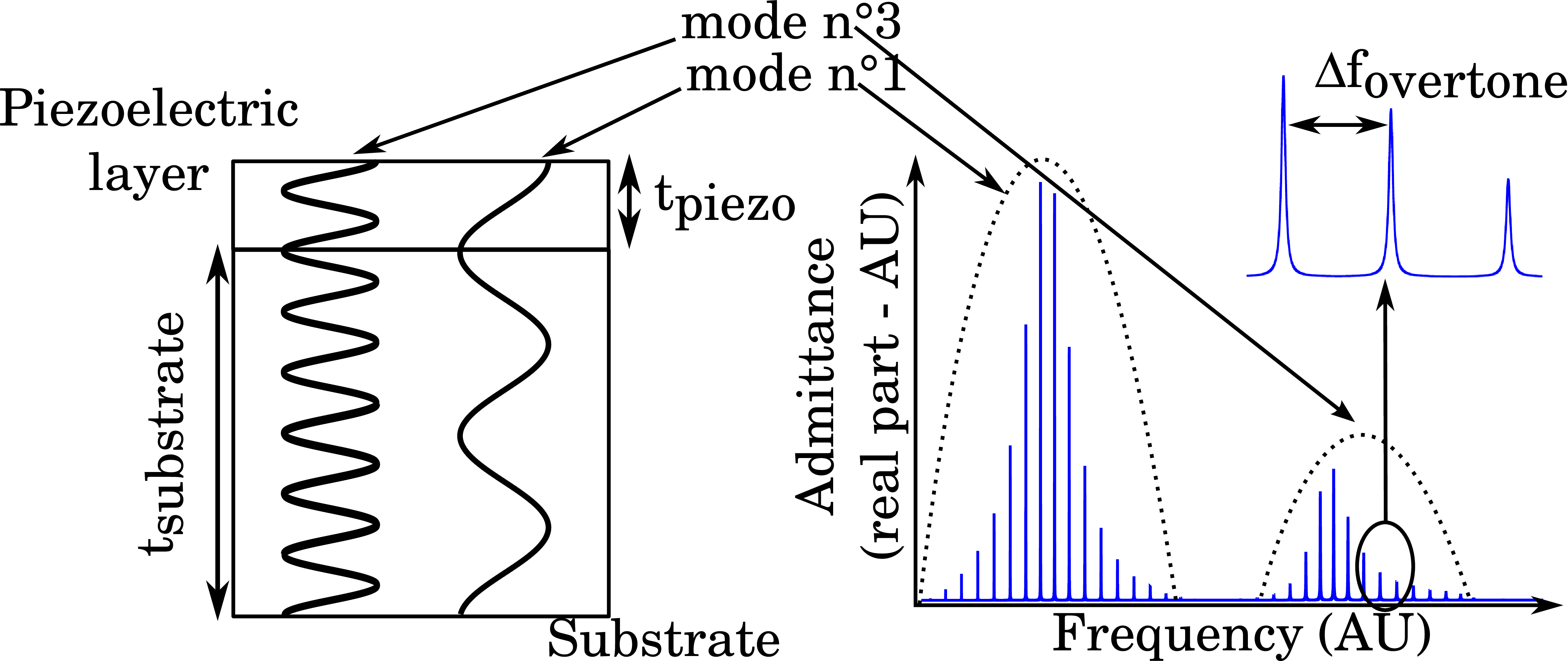}
		\caption{Principle of the HBAR (left) and global view of the real part of the
admittance (right).}
		\label{descri_hbar}
	\end{center}
\end{figure}

A HBAR device is a composite resonator including two layers: a thin piezoelectric layer 
(as a thin QCM) to generate the acoustic wave, and a low acoustic loss substrate used 
as a cavity to confine the resonances while supporting the thin piezoelectric film. 
This coupled resonator structure induces a complex admittance spectrum (Fig. \ref{descri_hbar})
with a series of narrow resonances whose amplitudes are modulated throughout the spectrum.

The envelope of the HBAR response is defined by the piezoelectric layer 
thickness while the frequency span between each narrow resonance is defined 
by the substrate layer thickness. Considering first only the piezoelectric layer
of thickness $t_{p}$, the resonance frequencies $f_{mode}(n)$ are related
to the acoustic velocity $c$ by  
\begin{equation}
f_{mode}(n) = n \times \frac{c}{2t_{p}}
\label{fmode}
\end{equation}
at which the envelope of the admittance is maximum since the piezoelectric thin
film pumps a maximum of energy in the substrate by inverse piezoelectric electromechanical
conversion.

Once the acoustic energy has been coupled to the substrate of thickness $t_{s}$, the 
frequency spacing $\Delta f_{overtone}$ between narrow resonances is given by
\begin{equation}
\Delta f_{overtone} \approx \frac{c}{2t_{s}}
\label{fharmo}
\end{equation}

This multitude of modes opens a unique perspective for exploiting the HBAR as
gravimetric transducer: wideband acoustic spectroscopy of the mechanical properties
of the adsorbed thin film. However, such an approach can only be exploited quantitatively
if the gravimetric sensitivity of each mode is known.

Two HBAR geometries are considered. A 3.8~\m m thick AlN piezoelectric 
thin film deposited on a 25.3~\m m-thick SiO$_2$ substrate only confines longitudinal waves 
exhibiting wavelengths ranging from 2 to 8~\m m when operating at frequencies in the
500~MHz to 5~GHz range. Since longitudinal waves are not appropriate for sensing in
liquid media (acoustic radiative losses), the second geometry combines {-- {\color{red} following
the IEEE 176-1987 (section 3.6) naming convention --} a lithium
niobate LiNbO$_3$ YXl/163$^o$ thin film (selected for its high coupling characteristics) over a
YXl/32$^o$ quartz substrate
(selected for its low acoustic losses characteristics and low temperature sensitivity): the 
very different technological processes induce thicker layers of 20~\m m \cite{masson2007dispersive} and 450~\m m 
respectively \cite{ballandras2011high,baron2011rf}. {\color{red}The latter device propagates 
pure shear waves and is hence compatible with the detection of compounds in liquid phase.}

\section{Modeling}

For modeling the HBAR resonator admittance and determining the gravimetric
sensitivity of the various overtones as boundary conditions are varied, 
a one dimension modeling software is used based on Boundary Element Modeling (BEM) 
\cite{reinhardt2003scattering,ballandras2005periodic}. The free parameters tuned
during the modeling process are the layer thicknesses and material properties,
while the gravimetric sensitivity is extracted from the application of Eq. \ref{sensi1}
when the resonance frequency is monitored as a function of adlayer geometrical
properties and most significantly its thickness $\Delta e$.

\subsection{Gravimetric sensitivity dependencies}

The study first focuses on the impact of the side of the HBAR selected as the sensitive surface. 
Although a practical consideration naturally hints at using the side opposite to the piezoelectric
layer coated with electrodes as the sensing area, the gravimetric sensitivity of both
sides of the HBAR (the exposed area of the piezoelectric layer or the substrate) will exhibit
different coupling with the adlayer and hence different gravimetric sensitivities (Fig. 
\ref{dessus_dessous}). Two adlayer mechanical properties are considered by selecting material
constants of silica or copper. The gravimetric sensitivities are calculated by considering an 
adsorbed thickness of 5~nm to remain in a perturbative assumption.

\begin{figure}[h!tb]
\begin{center}
\includegraphics[width=\linewidth]{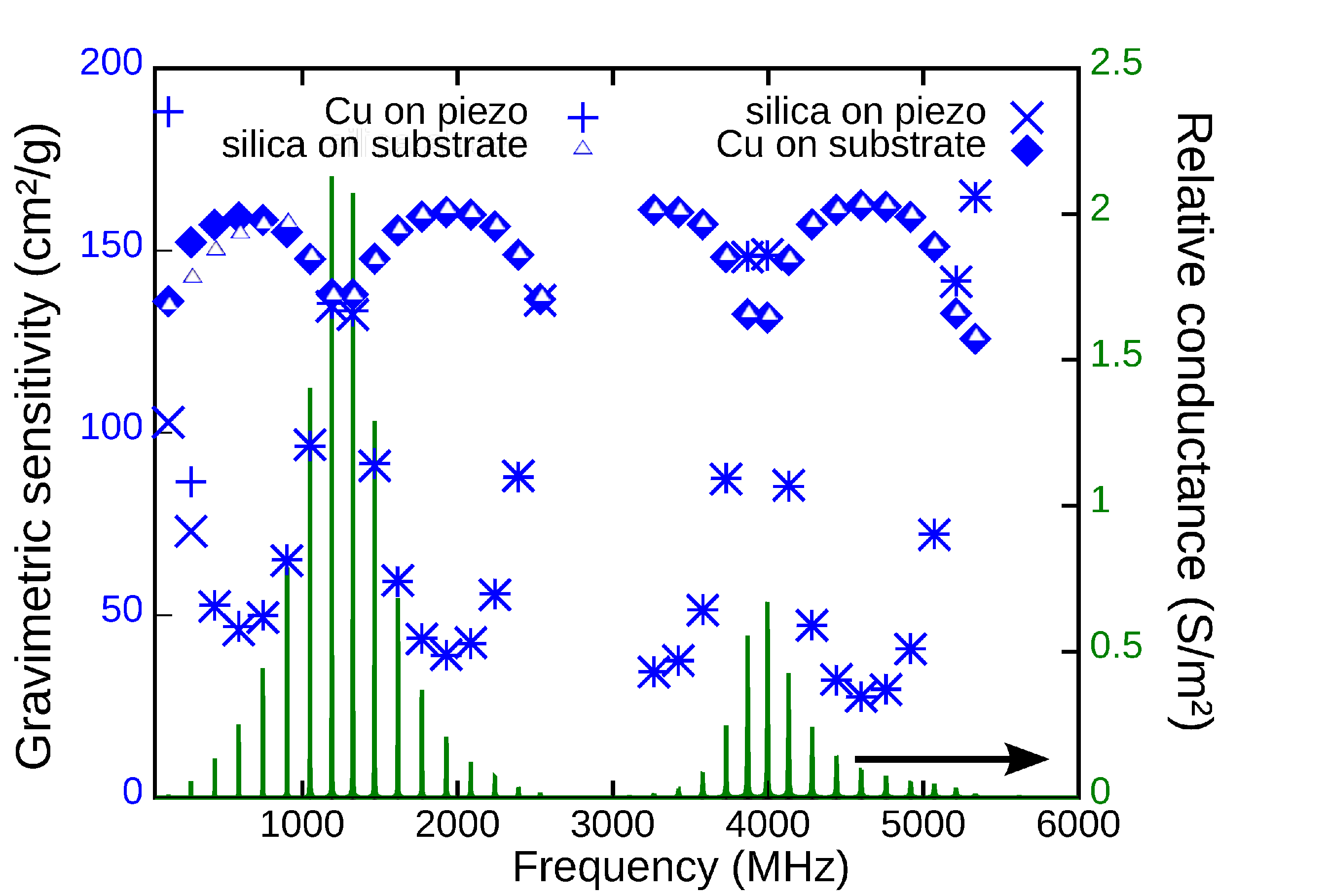}
\caption{Modeled HBAR admittance (solid line) and associated
gravimetric sensitivity using two materials (copper and silica) on the
top (piezoelectri layer) and bottom (substrate layer) sides
of the HBAR.}
\label{dessus_dessous}
\end{center}
\end{figure}

Both adlayer characteristics yield similar gravimetric sensitivities, complying with
the perturbation requirement of independence of the result with the thin additional
film material properties. However, the evolution of $S$ is radically different depending
on which side of the HBAR is considered. In the first case in which the adlayer coats the 
substrate side, the gravimetric sensitivity decreases when the admittance is maximized
and a tradeoff must be met between mode coupling and sensitivity: the gravimetric sensitivity
is maximized at resonance frequencies below and above the piezoelectric thin film resonance
frequency. The trend is opposite when coating the piezoelectric (top) side of the HBAR: in 
this case, both admittance and gravimetric sensitivity evole similarly. The gravimetric
sensitivities at the resonances of the piezoelectric thin film are the same whether the
coating is deposited on the bottom or top sides.

While the gravimetric sensitivity remains constant within 10\% when loading the substrate
side, it varies significantly -- in this case by a factor of 3 -- when loading the piezoelectric
layer side: hence, the probed modes must be carefully selected for maximizing both sensitivity
and signal to noise ratio through efficient electromechanical coupling.

\subsection{Thick film condition} 

In order to match experimental conditions, we shall from now on only consider an adlayer deposited
on the substrate side, opposite to the electrodes polarizing the piezoelectric thin film. This
strategy is selected so that packaging issues are only related to liquid confinement over the
HBAR sensing surface and no electrical insulation or shielding issues arise when operating with
compounds in liquid media.

Because on the wide range of operating frequencies, the perturbative assumption is hardly met
at the higher frequency range, and based on the previous work presented by Mansfeld in \cite{mansfeld2000theory}
we now focus on modeling the behaviour of thick absorbed films. ``Thick'' is defined as a film
exhibiting significant departure from the behaviour predicted by Sauerbrey. Considering a thick
film induces an uncertainty as to the definition of the initial condition when computing the
sensitivity. On the one hand, the sensitivity is defined as an infinitesimal frequency variation
due to an infinitesimal deposited mass: as such, the sensitivity is related to the derivate of the
frequency versus adsorbed layer thickness. This case is closely related to the one studied in
\cite{mansfeld2000theory} since a thick film acts as a gas absorbing layer and the sensitivity
of the transducer coated with the thick film is considered. On the other hand, if the initial condition 
is considered to be the layer-free transducer, then the sensitivity is computed as the frequency 
variation due to a thick absorbed layer, no longer complying with the derivate approximation only valid
for infinitesimal variations. The sensitivity computed by the latter approach is not only lower
than the sensitivity derived from the derivate approach, but the thickness at which the sensitivity
is maximized is not the same depending on the selected approach due to the curvature of the
frequency v.s thickness curve, as shown in Fig. \ref{diffs}. Such conditions match our experimental
assesment of the sensitivity by electro-depositing copper layers on the bare HBAR surface up to thicknesses
matching the wavelength. In the following text, we consider the former approach as a thin film approach, even 
though we are considering a small increase of an already thick layer, while the latter will be called 
the thick film approach,

\begin{figure}[h!tb]
\includegraphics[width=\linewidth]{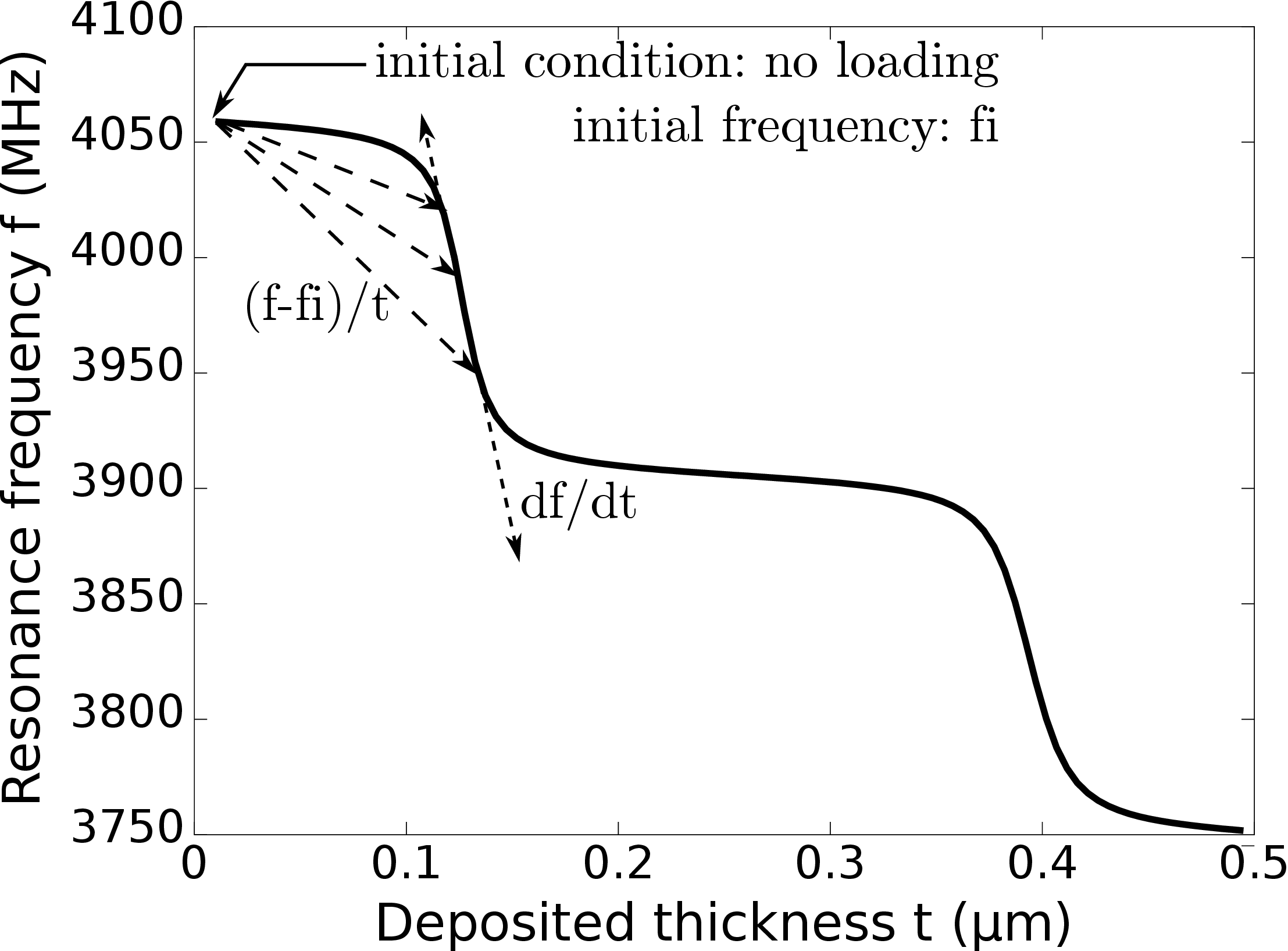}
\caption{Typical curve exhibiting the evolution of the resonance frequency of one mode of the
HBAR as a function of absorbed layer thickness: the sharp rise in the frequency v.s thickness
slope is observed for deposited thicknesses equal to multiples of the quarter wavelength. The
frequency variation due to an absorbed layer thickness $t$ depends on whether the initial
condition is considered to be the bare transducer or the transducer already coated with a thick
film. The latter approach always yields a larger estimate of the sensitivity than the former,
as shown by the dotted lines representing the local slope of the frequency v.s thickness
curve.}
\label{diffs}
\end{figure}

Departure from the perturbative assumption is considered by modeling an adlayer
thickness of the same order than the wavelength. The results in the thick film 
approach, for two working frequencies, 1324 and 4000~MHz corresponding to wavelengths of 
2.1 and 0.68~\m m respectively,
are presented in Fig. \ref{S(e)}. $S$ is calculated for thicknesses of an adlayer, assumed to
meet the material properties of copper, ranging from 50~nm to 2.5~\m m. Both overtones
exhibit oscillating gravimetric sensitivities as a function of adlayer thickness following the
initial drop, with a period dependent on the overtone wavelength, yet the asymptotic sensitivity
value remains the same at about 60~cm$^2$/g.

\begin{figure}[h!tb]
\includegraphics[width=\linewidth]{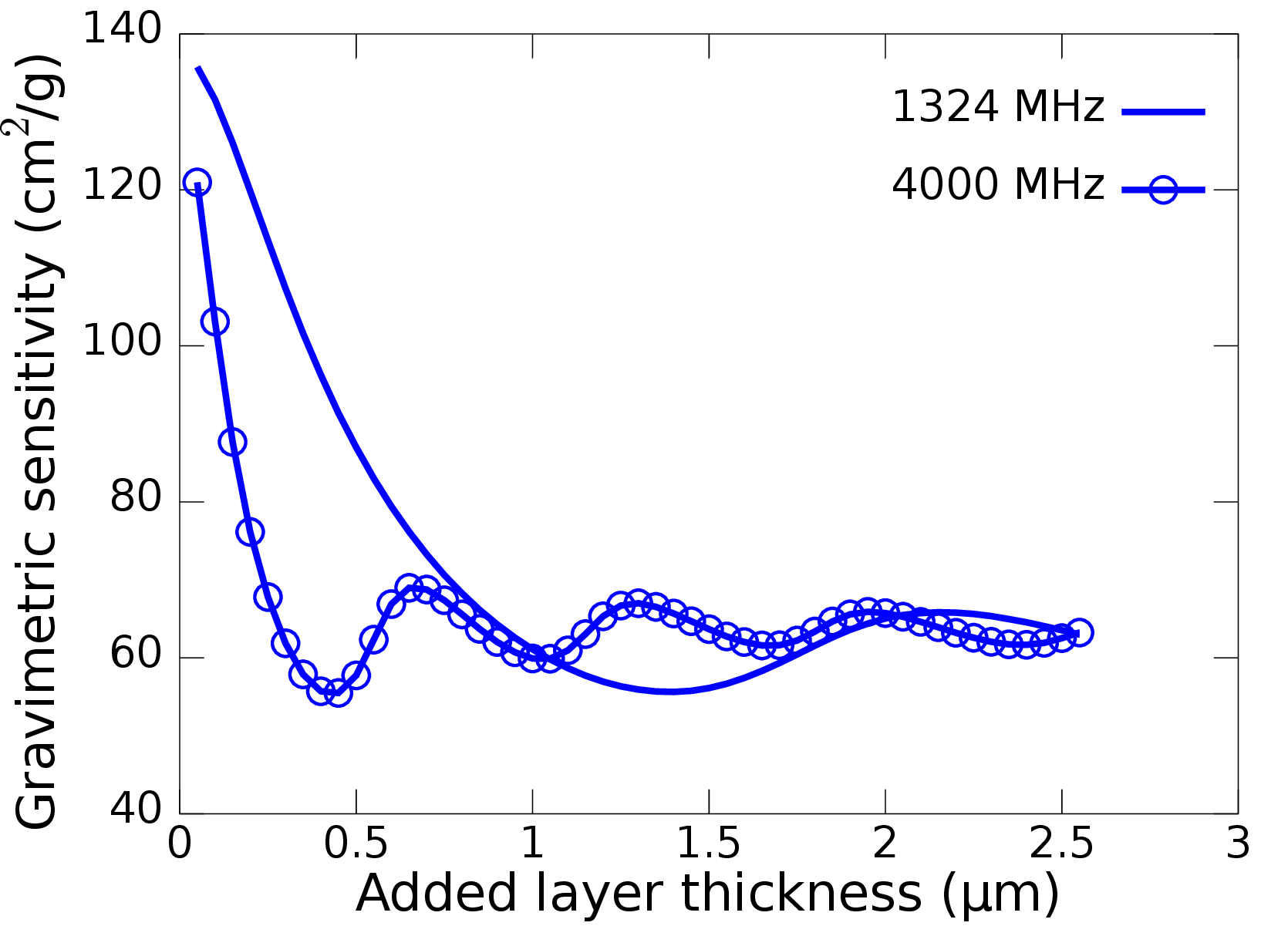}
\caption{Gravimetric sensitivity calculated as a function of the
thickness of the copper adlayer for 1324 (solid line) and 
4000~MHz (circles) resonance frequencies.}
\label{S(e)}
\end{figure}

The same analysis in the thin film approach provides a clearer view of the resonant
confinement of the acoustic energy in the thick absorbed film, as shown in Fig. \ref{S(t)}.
The frequency v.s thickness results are the same than those shown in Fig. \ref{S(e)} but
here the initial state for computing the sensitivity value is selected as the infinitesimaly
thinner layer, hence compatible with the derivate of the frequency v.s thickness computation.
Not only are the thicknesses at which sensitivity is maximum 
closely equal to multiples of the wavelength, 
but the actual sensitivity values remain close to the thin film value at
odd multiples of the half wavelength -- 120~cm$^2$/g -- as opposed to the thick film approach in which the
sensitivity remained consistently lower than the perturbative layer sensitivity.

\begin{figure}[h!tb]
\includegraphics[width=\linewidth]{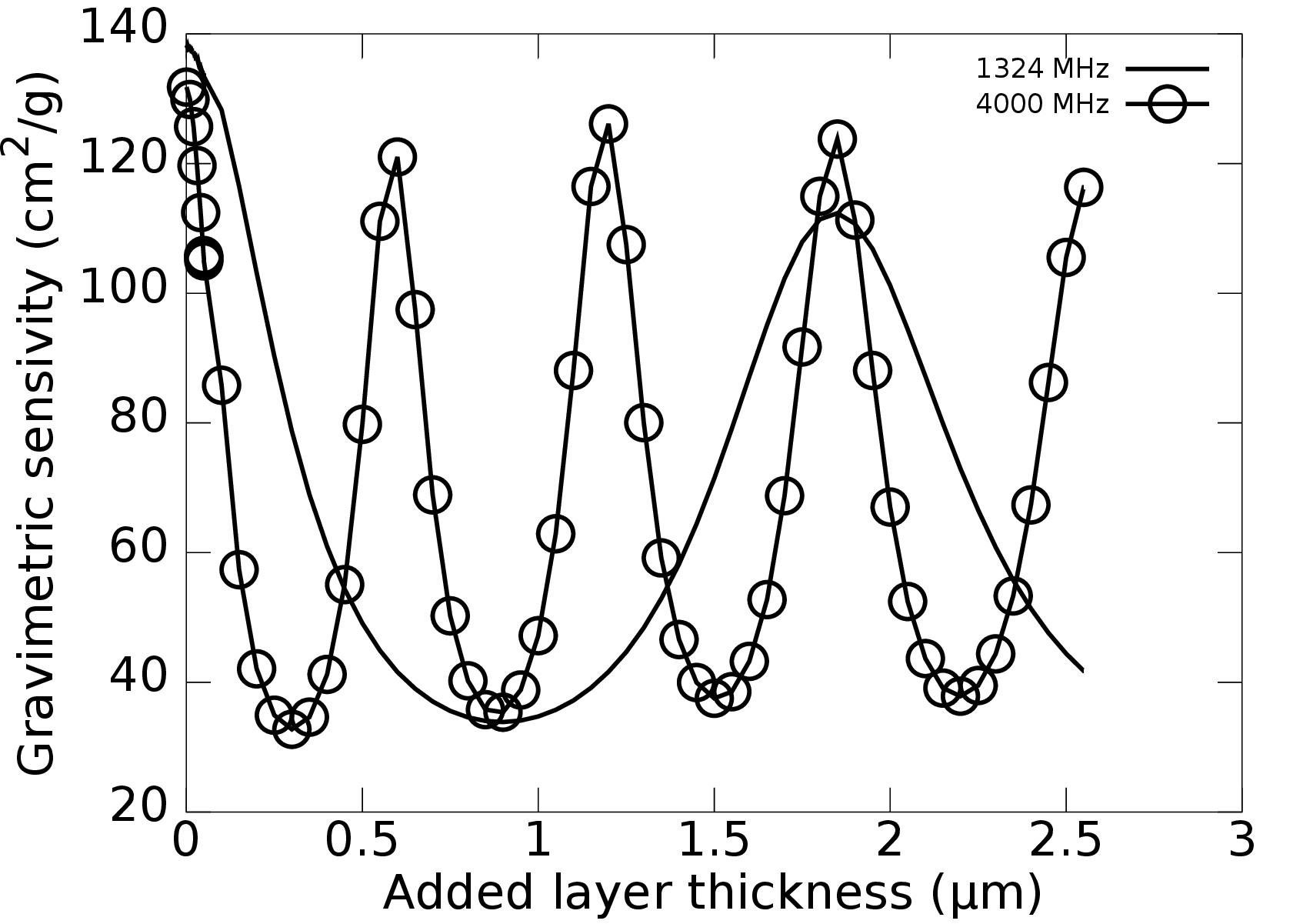}
\caption{Thin film analysis of the gravimetric sensitivity using the same simulation 
results than those exhibited in Fig. \ref{S(e)}.}
\label{S(t)}
\end{figure}

These results indicate that various overtones react differently to an adlayer of
varying thickness due to the evolution of the energy distribution between the three
layers -- adlayer, piezoelectric thin film and substrate -- in a coupled resonator
context, making the wideband acoustic spectroscopy analysis non-trivial. 
An optimum operating frequency can be selected if the adlayer thickness is fixed and
known in order to maximize $S$: such a conclusion was already reached in a previous
analysis \cite{mansfeld2000theory}.
However, Mansfeld \cite{mansfeld2000theory} determined theoretically and experimentally 
that the adlayer thickness maximizing $S$ would be $\lambda/4$: this conclusion is not 
validated in the present case. To investigate the cause of the differences, several kinds of 
adsorbed material (Tab. \ref{table_imp}) used as perturbative layer are considered to 
assess the dependence of this
conclusion with adlayer properties (Fig. \ref{S_imp}).  

\begin{figure}[h!tb]
\begin{center}
\includegraphics[width=\linewidth]{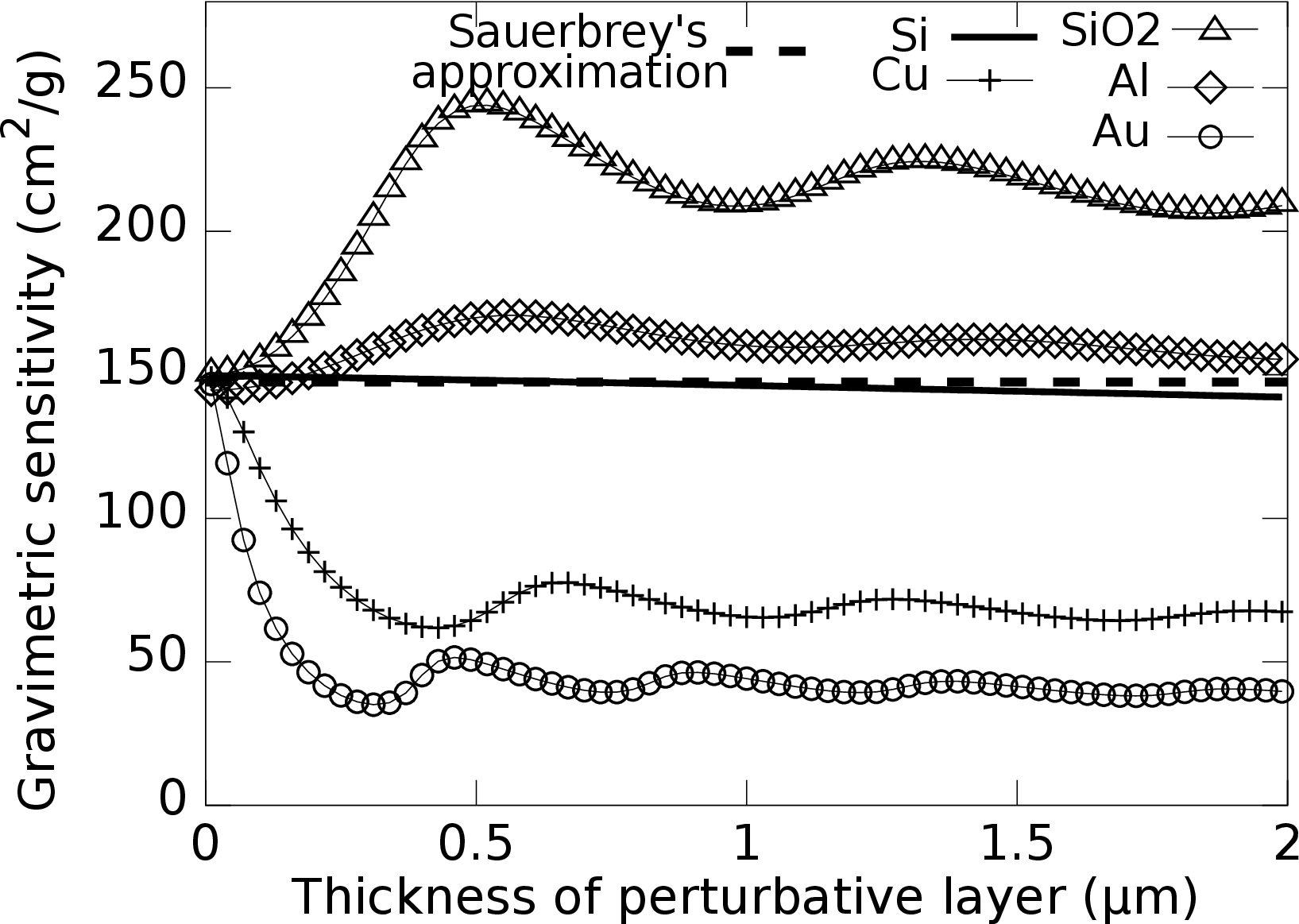}
\caption{Gravimetric sensitivity as a function of adlayer thickness,
calculated for a working frequency of 4~GHz. Silica, aluminum, chromium, silicon 
and gold are considered as perturbative layer materials when computing the the 
gravimetric sensitivity. The dashed line corresponds to the gravimetric 
sensitivity calculated by Sauerbrey's approximation, considering a silicon 
resonator with a thickness of 29.1~\m m, equal to the global thickness of the 
simulated HBAR.}
\label{S_imp}
\end{center}
\end{figure}

The validity of the approach is assessed by first considering a silicon adlayer --
the same material the HBAR is made off -- and
checking that the resulting sensitivity is indeed equal to the value predicted
by the Sauerbrey perturbation theory (Fig. \ref{S_imp}, dashed line and solid line)
and independent of the adlayer thickness. For other adlayer materials (copper and 
gold) with an acoustic impedance higher than that of the silicon substrate, $S$ 
decreases with increasing thickness as was previously observed in Fig. \ref{S(e)}.
For materials (silica and aluminum) with lower acoustic impedance than that of the
substrate, the gravimetric sensitivity increases when the thickness increases. 

\begin{table}[h!tb]
\caption{Acoustic impedances of the materials used for gravimetric sensitivity determination of 
a silicon HBAR.}
\begin{tabular}{|c||c|c|c|}
	\hline
	\multirow{2}{*}{{materials}}& ~~~~~~$Z_{ac}$ ~~~~~~  & ~~~ velocity ~~~ \\
	& (MRayl)   & (m/s)   \\
	\hline
	Silica (SiO$_2$)   & 13 & 5740 \\
	\hline
	Aluminum (Al)   & 14 & 5018 \\
	\hline
	{\bf Silicon (Si)   } & {\bf17} & 7483 \\
	\hline
	Copper (Cu)   & 24 & 2728 \\
	\hline
	Gold (Au)    & 29 & 1480 \\
	\hline
	Aluminum nitride (AlN) & 30 & 11500\\
	\hline
	YAG    & 36 & 7801 \\
	\hline
\end{tabular}
\label{table_imp}
\end{table}

These results demonstrate that the maximum of the gravimetric sensitivity depends on 
the relative acoustic impedances and adlayer thickness to wavelength ratio.
Furthermore, an increase of the gravimetric sensitivity can be obtained 
with an adsorbed material with acoustic impedance lower than that of the substrate as
is classically known from the Love wave configuration: such an approach will
be discussed in section \ref{V}.

\subsection{Comparison with Mansfeld's theory}

Although numerical constants are not provided in \cite{mansfeld2000theory} for a direct
comparison with these results, their use of YAG as a high acoustic impedance substrate 
\cite{mezeix2006comparison}, exhibiting a high acoustic velocity, as the HBAR substrate,
and the organic layer acting as the adlayer, hints at a case in which a low impedance coating 
is deposited over a high impedance substrate. Such a stack matches the qualitative behaviour 
identified by our numerical simulation. The quantitative assesment of the layer
thickness maximizing the gravimetric sensitivity however requires an in-depth
analysis of the sensitivity dependence with material properties. Such considerations are
demonstrated in Fig. \ref{S_mansfeld} which exhibits the acoustic wavelength (normalized to 
the layer thickness) at which the gravimetric sensitivity is maximized, as a function
of the adlayer acoustic impedance. 
The gravimetric sensitivity is calculed using the thin film
approach to be comparable with \cite{mansfeld2000theory} in which the resonant frequency variation
 is recorded for an infinitesimal thickness variation of the adlayer due to gas adosorption.
The results presented here consider an adlayer material with a 
constant Young's modulus (13~GPa) and various densities and Poisson coefficients. The elastic 
constants (C$_{11}$, C$_{12}$, C$_{66}$) of the material are calculated for each 
density and Poisson coefficent values.
Maximizing the sensitivity for a $\lambda/4$ thickness of the adlayer is consistent in some cases
which present a low Poisson coefficient (less than 0.2) and different acoustic impedance of the 
adsorbed material. The BEM approach used here takes in account 
all the elastic constants of the materials and so exhibits more rigorous results than the 
analytical approach.

\begin{figure}[h!tb]
	\begin{center}
		\includegraphics[width=\linewidth]{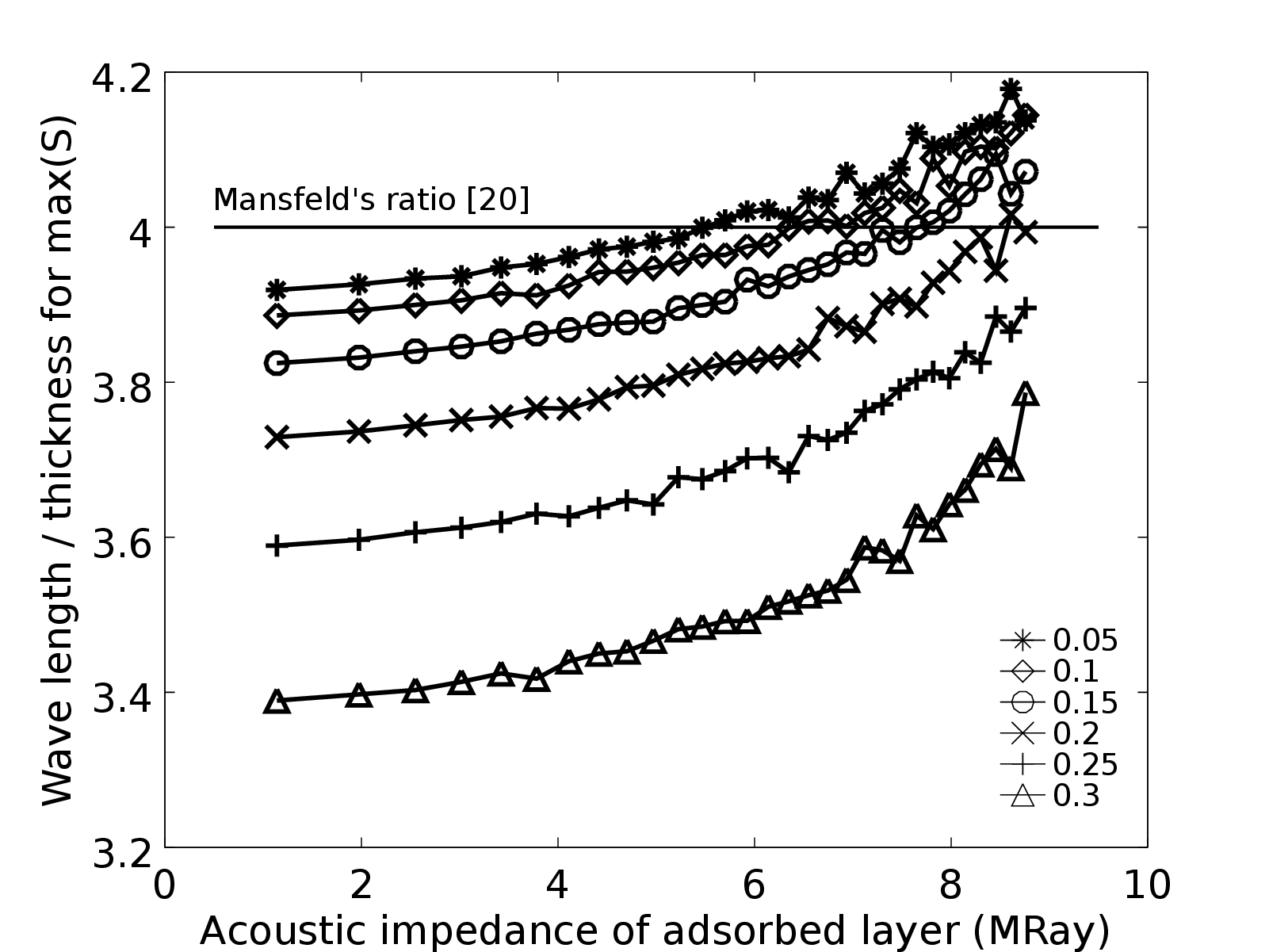}
		\caption{Ratio of the acoustic wavelength in the adlayer to the adlayer thickness
maximizing the gravimetric sensitivity. The solid line indicates the value cited by Mansfeld
when considering an organic layer over a YAG substrate, the marked lines are the result of our
simulation for varying acoustic impedance adlayers and using different poisson coefficients
 over a silicon HBAR.}
		\label{S_mansfeld}
	\end{center}
\end{figure}

The gravimetric sensitivity dependence with overtone number (and hence wavelength) and 
material property of the adlayer has been investigated through simulation, demonstrating 
a non-trivial link between these relations. A low impedance adlayer is predicted to 
magnify the gravimetric sensitivity. Moreover, results cited earlier in the literature
could be modeled in detail during these investigations, whose results will now be 
confronted to experimental results.

\section{Experimental results}

Experimental assessment of the gravimetric sensitivity of HBARs is performed
in two distinct steps: on the one hand the irreversible deposition of thin
copper films in a cleanroom environment by sputtering, and on the other hand the reversible
electrodeposition of copper in a wet environment. All depositions are performed
on the substrate side of the HBAR, opposite to the electrodes deposited on the
piezoelectric thin film. The admittance of the HBAR is monitored by a network analyzer,
either after each deposition step in the case of the sputtering, or continuously during
the electrochemical oxydation and reduction cycles. Since part of these experiments
will be performed in a wet environment, only the lithium niobate over quartz HBAR propagating
pure shear waves is considered.

The HBAR is characterized at four different frequency ranges (280-310 ; 
410-440 ;
670-700 ; 800-830~MHz). Each frequency range presents about seven resonances. 
As shown on table \ref{table_sensi}, the gravimetric sensitivity for each 
resonance is calculated by considering the initial resonant frequency as the 
resonance frequency obtained with the previous adlayer thickness (thin film approach). 
The acoustic wavelengths for each frequency range are close, so only 
the mean value of the gravimetric sensitivities is presented and all thicknesses
are normalized to the acoustic wavelength in the adlayer (Fig. \ref{exp_vs_simu1}).

\begin{table}[h!tb]
\caption{Experimental gravimetric sensitivity mean value
calculated for each frequency range and for each deposited copper thickness.}
	\footnotesize 
\hspace*{-0.6cm}\begin{tabular}{|c|c|c|c|c|}
		\hline
		Deposited & \multicolumn{4}{c|}{Mean of gravimetric sensitivity (cm$^2$/g)} \\
		\cline{2-5}
		thickness (nm)& 280-310~MHz & 410-440~MHz & 670-700~MHz & 800-830~MHz \\ 
		\hline
		  196 & 10.5 & 7.2 & 4.5  & 5.1 \\
		  381 & 7.7 & 5.5 & 3.5  & 3.7 \\
		  541 & 7.0 & 5.0 & 3.3  & 3.4 \\
		  726 & 6.3 & 4.5 & 3.1  & 3.4 \\
		  891 & 5.8 & 4.2 & 3.2  & 3.9 \\
		  1099 & 5.2 & 3.8 & 3.5  & 4.6 \\
		  1299 & 4.8 & 3.6 & 4.2  & 4.8 \\
		  1514 & 4.4 & 3.4 & 4.6  & 4.5\\
		\hline
		
	\end{tabular}
	\label{table_sensi}
\end{table}

Both experiemental and modeled (Fig. \ref{exp_vs_simu1}) 
dependences of the gravimetric sensitivity
with the adlayer thickness hint at a starting value of about 10~cm$^2$/g and a
secondary maximum.
{\color{red}The discrepancy between the modeled and experimental results, yielding 
different adlayer thicknesses maximizing the sensitivity, is attributed to the 
use of bulk material constants which might not appropriately represent
the thin copper film properties.
}

\begin{figure}[h!tb]
\begin{center}
\includegraphics[width=\linewidth]{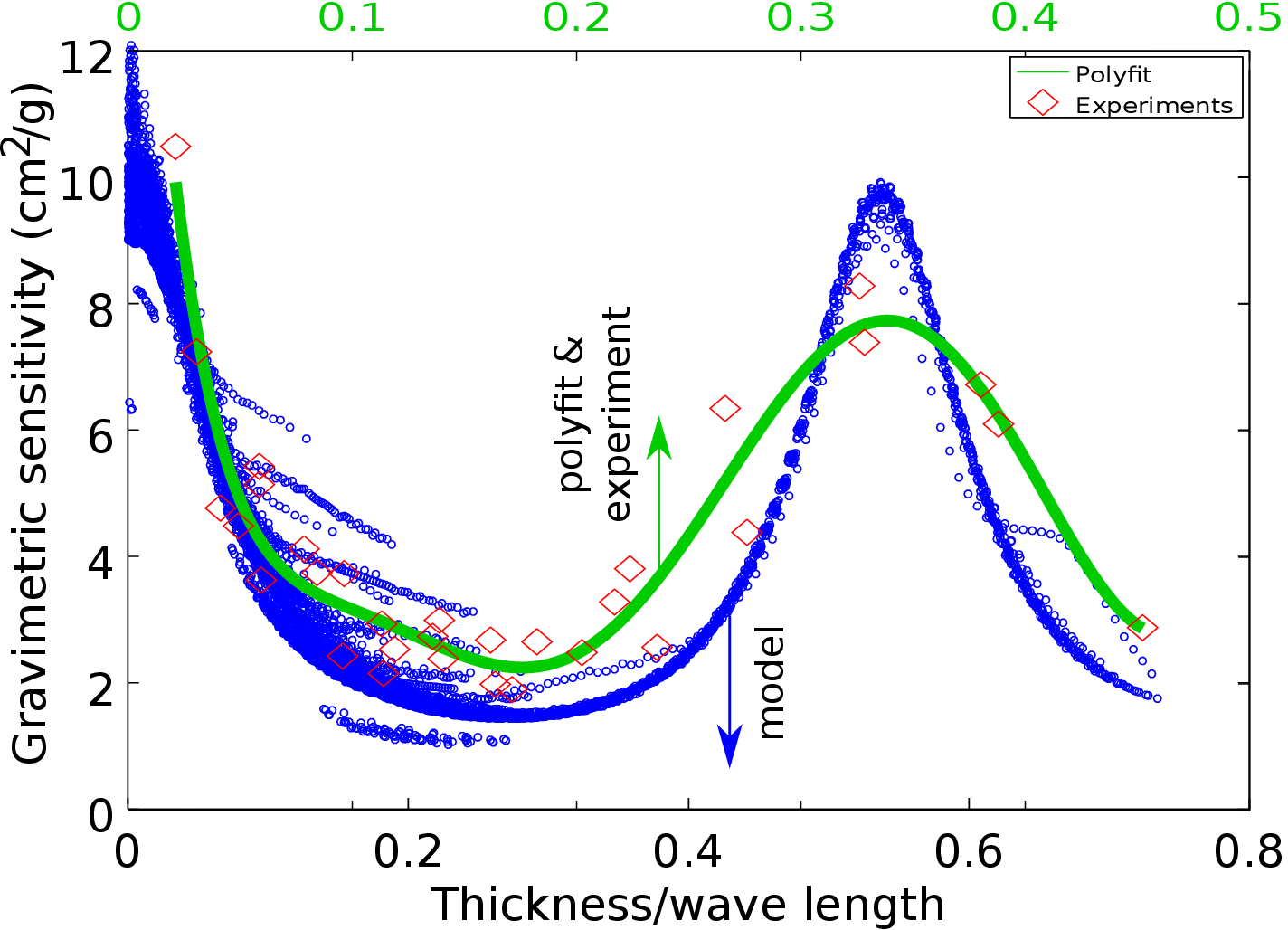}
\caption{Mean value of the measured gravimetric sensitivity for
each frequency range (diamonds) as a function of the deposited thickness 
divided by the acoustic wavelength, and polynomial fit (solid line) as a guide
for the eye. Circles: gravimetric sensitivity estimates resulting from modeling the 
lithium niobate over quartz HBAR stack considered in the experimental section: 
the different curves are associated with different overtones. Notice that the two
charts do not share the same abscissa: the experimental data abscissa is given on
top, the model abscissa is given on the bottom. The gravimetric sensitivity (ordinate)
is properly modeled and shared by the two charts.}
\label{exp_vs_simu1}
\end{center}
\end{figure}

An alternative to cleanroom sputtering of copper is the use of electrochemical 
deposition on the sensing surface of an HBAR. This approach, already used to 
characterize the gravimetric sensitivity of QCM
\cite{friedt2003simultaneous}, SAW \cite{avs} and HBAR \cite{rabus2012eight} devices, is
attractive because it is reversible (allowing for multiple cycles for assessing the
reproducibility of the result) and operates in liquid phase, hence being more
representative of the behaviour of the sensor used for detecting compounds in 
aqueous solutions (e.g. biosensing). This method is only usable with devices
propagating pure shear waves due to viscoelastic coupling of the propagating longitudinal
waves. The chemical reaction is driven by a custom-made potentiostat included in
the embedded electronics \cite{rabus2013high} designed to probe simultaneously multiple
overtones of the HBAR. This electronics provides a measurement rate large enough to be 
compatible with the reaction kinetics. 

The gravimetric sensitivity of an overtone at 327~MHz of the lithium niobate/quartz
HBAR is investigated: electrochemical deposition provides an independent estimate
of the adlayer mass $m_{Cu}$ through Eq. \ref{sensi1}, assuming a 100\% yield, by 
considering the number of electrons involved in the reduction process as the
integral of the current $i(t)$ flowing through the working electrode
\begin{equation}
m_{Cu} = \frac{M_{Cu}\times\Sigma i(t) \delta t}{N_A \times e \times n_e} 
\label{masse}
\end{equation}
where $M_{Cu}$ is the molar weight ($g/mol$) of the adlayer, $\Sigma i(t) \delta t$ the 
number of charges transferred during electro-deposition, considering that the charge of 
one mole of electrons ($N_A$) is $96440$~C, and $n_e$ the number of electrons transfered 
during the redox 
reaction (Eq. \ref{formule_reaction})
\begin{equation}
Cu^{2+} + 2e^- \leftrightarrow Cu
\label{formule_reaction}
\end{equation} 

Fig. \ref{S_electro} exhibits the gravimetric sensitivity measured
using the electro-deposition approach and the modeling of the used HBAR, both 
considered at the same working frequency. This working frequency is fixed and used as the inital
resonant frequency (thick film approach). 
The thin film approach for calculating the gravimetric sensitivity could not
be used in this case due the experimental set up which does not allow to have the resonant frequency 
between each adlayer thickness. Knowing the area $A$ of the
sensing side of the HBAR over which the electrochemical reaction occurs, Eq. \ref{sensi1} and \ref{masse}
allow for estimating the deposited thickness. Hence, the gravimetric sensitivity is
plotted as a function of the deposited thickness. Experiment matches the modeled sensitivities
for adlayer thicknesses above 1.2~\m m. Below this value, the calculated sensitivity 
is 3 to 6 times higher than the model results. The main cause of divergence of the two
curves for thin adlayers is attributed to the inhomogeneous deposition which starts
at the center of the HBAR sensing area. In such cases, the estimated adlayer
thickness $\Delta e=m_{Cu}/A$ is under-estimated since $A$ is overestimated when
using the geometrical area, and the experimental sensitivity is hence over-estimated.

\begin{figure}[h!tb]
\begin{center}
\includegraphics[width=\linewidth]{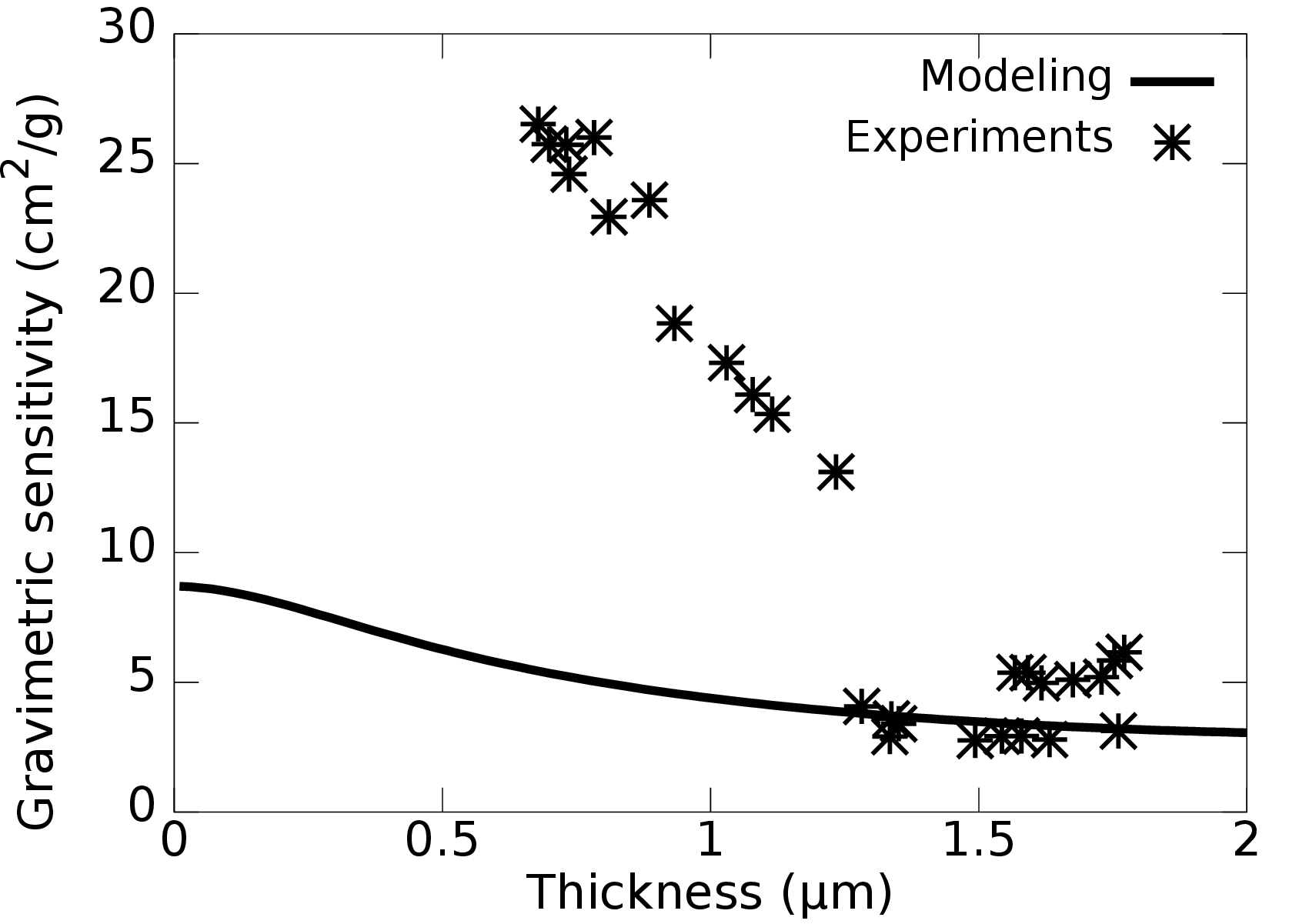}
\end{center}
\caption{Gravimetric sensitivity measured (stars) and modeled (solid line) for a 
lithium niobate over quartz HBAR operating at 327~MHz, as a function of the 
electrochemically deposited layer thickness.}
\label{S_electro}
\end{figure}

Based on these considerations, the HBAR geometries considered so far exhibit
sensitivities consistent with those of bulk QCMs and hence 10 to 20 times lower
than SAW devices operating in the hundreds of MHz range. However, the low-acoustic
impedance adlayer has been shown to increase the gravimetric sensitivity, so we 
consider whether an additional stack of material over the HBAR might bring some 
gravimetric sensitivity improvement, aiming at the hundreds of cm$^2$/g range classically
found for Love-mode SAW devices \cite{avs}.

\section{Gravimetric sensitivity improvement}\label{V}

Two ways to improve the gravimetric sensitivity have been theoretically explored.
Since the Sauerbrey gravimetric sensitivity depends on the 
working frequency which depends on the thickness of the QCM, reducing the 
overall sensor thickness will be considered. Based on this 
idea, the gravimetric sensitivity is calculated for lithium niobate over quartz
HBARs when varying the substrate thickness from 56.25 to 450~\m m (Tab. \ref{table_S_quartz}). 

\begin{table}[h!tb]
\caption{Calculated gravimetric sensitivity for different thicknesses of quartz
	substrate. Frequency ranges and the number of probed modes are also 
	presented.}
\label{table_S_quartz}
	\begin{tabular}{|c||c|c|c|c|c|c|c|c|}
		\hline
		Frequency & \multicolumn{4}{c|}{\multirow{2}{*}{50 - 150}}&
		\multicolumn{4}{c|}{\multirow{2}{*}{ 300 - 550}} \\
		range (MHz) & \multicolumn{4}{c|}{} & \multicolumn{4}{c|}{}\\
		\hline
		Quartz & \multirow{2}{*}{450}& \multirow{2}{*}{225}&
		\multirow{2}{*}{112.5}& \multirow{2}{*}{56.25}& \multirow{2}{*}{450}& 
		\multirow{2}{*}{225}& \multirow{2}{*}{112.5}& \multirow{2}{*}{56.25} \\
		thickness ($\mu$m) & & & & & & & & \\
		\hline
		number of& \multirow{2}{*}{53}& \multirow{2}{*}{28}&
		\multirow{2}{*}{14}& \multirow{2}{*}{7}& \multirow{2}{*}{51}& 
		\multirow{2}{*}{26}& \multirow{2}{*}{13}& \multirow{2}{*}{8} \\
		probed modes & & & & & & & & \\
		\hline
		avg. sensitivity & \multirow{2}{*}{9}& \multirow{2}{*}{18}& \multirow{2}{*}{37}& \multirow{2}{*}{80}& \multirow{2}{*}{8}& \multirow{2}{*}{16}& \multirow{2}{*}{31}& \multirow{2}{*}{59} \\ 
		(cm$^2$/g)   & & & & & & & & \\
		\hline
		theoretical sens.& \multirow{2}{*}{8}& \multirow{2}{*}{17}& 
		\multirow{2}{*}{34}& \multirow{2}{*}{67}& \multirow{2}{*}{8}& 
		\multirow{2}{*}{17}& \multirow{2}{*}{33}& \multirow{2}{*}{67} \\ 
		(Sauerbrey. cm$^2$/g)   & & & & & & & & \\
		\hline 
	\end{tabular}
\end{table}

Although this approach trivially scales the sensitivity as the substrate 
thinning ratio,
closely matching the Sauerbrey equation prediction, the transducer ruggedness is impacted
and the solution is not satisfactory in reaching disadvantages of FBARs.

A second investigated way of improvement is adding a well-chosen material on 
the sensitive surface of the HBAR. The gravimetric sensitivity depends on the 
impedance of the deposited materials (Fig. \ref{S_imp}). Following a strategy 
proven in the case of the Love-mode SAW transducer, an additional layer is designed 
to confine the acoustic wave near the sensing surface to improve the gravimetric sensitivity. 
The efficiency of this approach is assessed by modeling the sensitivity of an AlN
over silicon HBAR, coated with an additional layer of silicon oxide. In this calculation,
the gravimetric sensitivity is calculated by considering a 5~nm-thick copper adlayer
on the silicon oxide (Fig. \ref{S_SiO2_Cu}). 

\begin{figure}[h!tb]
\begin{center}
\includegraphics[width=\linewidth]{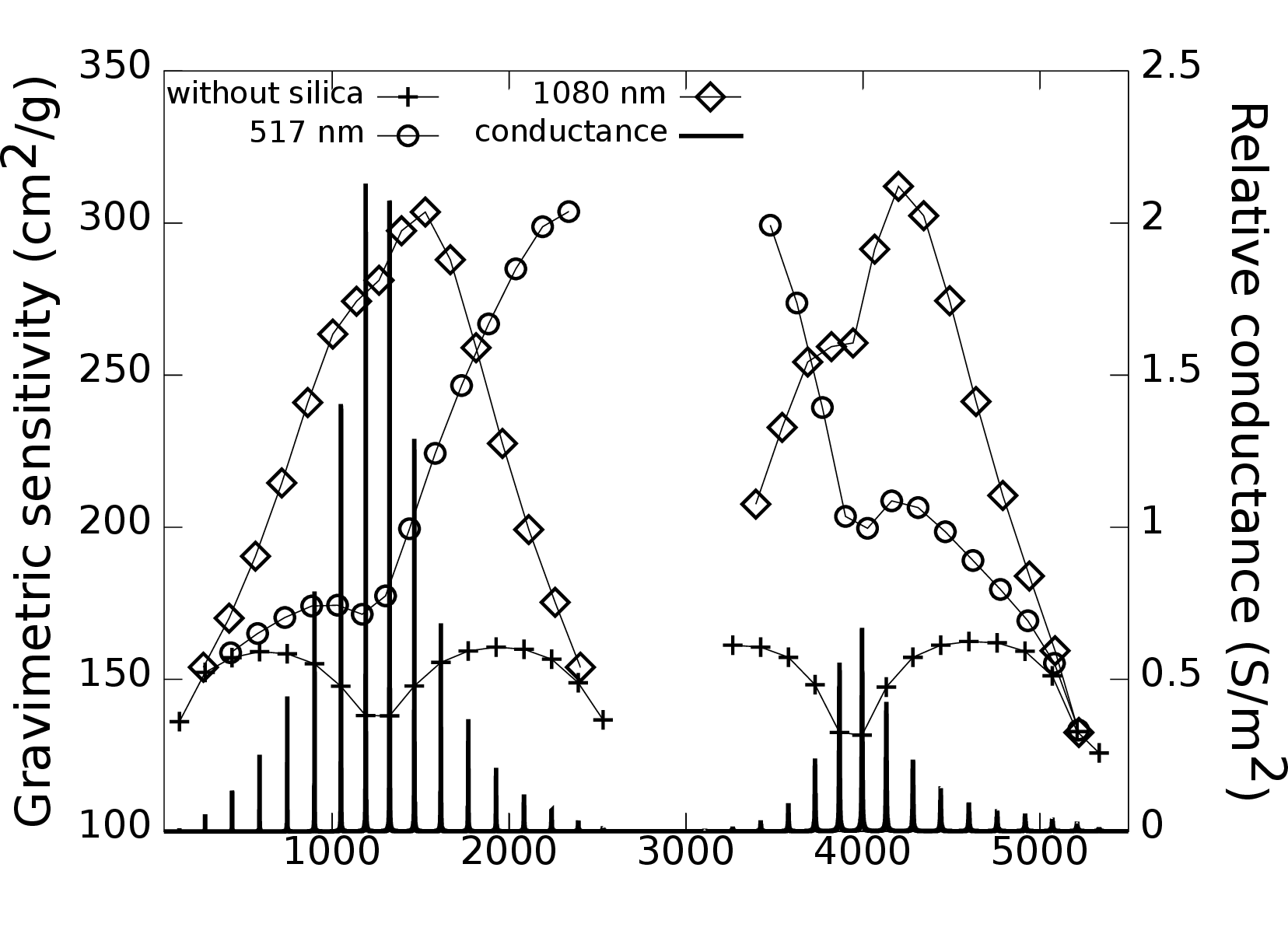}
\caption{Gravimetric sensitivity calculated for an AlN over Si
HBAR with (lines with markers) and without (solid line) 
silicon oxide layer.}
\label{S_SiO2_Cu}
\end{center}
\end{figure}

Different thicknesses of the silicon dioxide acoustic field confinement layer
are considered (Fig. \ref{S_SiO2_Cu}): the affected overtone varies as a function
of the silicon dioxide layer thickness, but in all cases a dramatic sensitivity enhancement is
observed, with a doubling of the sensitivity with respect to the bare device.

\section{Conclusions}

The gravimetric sensitivity of composite HBAR resonators has been studied to 
determine their potential as direct detection sensors. Two architectures,
aluminum nitride over silicon and lithium niobate over quartz, are considered
as complementary since the former exhibits high sensitivity {\color{red} -- of
the same order of magnitude as those found for 125~MHz Love mode SAW devices --} but propagates
longitudinal waves incompatible {\color{red} with} sensing compounds in liquid phase, while the
latter propagates pure shear waves {\color{red} yet only exhibits sensitivity with
values around those exhibits by radiofrequency bulk acoustic resonators -- typically
10 times lower than the Love-mode value}. The multimode spectral characteristics
of these transducers is considered best suited for wideband acoustic spectroscopy
of adsorbed layers. However, the complex dependence of the gravimetric sensitivity
of the various overtones yields non-trivial analysis considerations requiring
accurate acoustic behaviour modeling of the coupled acoustic fields in the various 
layers. The poor gravimetric sensitivity of the bare device is theoretically
improved by adding a low-acoustic impedance layer on the sensing area following
a strategy reminiscent of the Love mode guided SAW device. Working on the
electrode-free side of the HBAR solves the classical packaging issue of SAW devices
since no structure needs to be located on the acoustic path while electrodes
are prevented from being in contact with the medium containing the analyte being
investigated.

\section*{Acknowledgements}

This work was supported by the French RENATECH network and its FEMTO-ST 
technological facility. Part of this work was funded by the French DGA through the 
ROHLEX grant and a Defense PhD funding, as well as by the European LOVEFOOD project
(FP7-ICT-2011.3.2 grant).

\bibliographystyle{IEEEtran}
\bibliography{biblio}
\end{document}